\documentclass[pra,aps,showpacs,twocolumn,amsmath,amssymb,nofootinbib,10pt]{revtex4-1}

\usepackage{graphicx,dcolumn,bm}
\usepackage{amsmath}
\usepackage{amsthm}
\usepackage{amsfonts}
\usepackage{mathrsfs}
\usepackage[usenames,dvipsnames]{xcolor}
\usepackage[colorlinks=true,citecolor=MidnightBlue,linkcolor=MidnightBlue,urlcolor=MidnightBlue]{hyperref}

\usepackage{bbm}

\newcommand{\average}[1]{\langle #1 \rangle}

\newcommand{\fr}{\mathbf{r}}

\newcommand{\ket}[1]{|#1\rangle}
\newcommand{\bra}[1]{\langle #1|}
\renewcommand{\vec}[1]{{\bf #1}}

\newcommand {\startbild}{\begin{figure}}
\newcommand {\stopbild}{\end{figure}}
\newcommand {\staf}{\begin{equation}}
\newcommand {\stof}{\end{equation}}
\newcommand {\staffeld}{\begin{eqnarray}}
\newcommand {\stoffeld}{\end{eqnarray}}
\newcommand {\staa}{\begin{align}}
\newcommand {\stoa}{\end{align}}

\newcommand{\cref}[1]{Chapter~\ref{#1}}     

\begin{document}

\title{Simulating superradiance from higher-order-intensity-correlation measurements: Single atoms}
 
 \author{R.~Wiegner$^{1}$}
\affiliation{$^{1}$Institut f\"{u}r Optik, Information und Photonik, Universit\"{a}t Erlangen-N\"{u}rnberg, 91058 Erlangen, Germany}
\affiliation{$^{2}$Erlangen Graduate School in Advanced Optical Technologies (SAOT), Universit\"at Erlangen-N\"urnberg, 91052 Erlangen, Germany}

\author{S.~Oppel$^{1,2}$}
\affiliation{$^{1}$Institut f\"{u}r Optik, Information und Photonik, Universit\"{a}t Erlangen-N\"{u}rnberg, 91058 Erlangen, Germany}
\affiliation{$^{2}$Erlangen Graduate School in Advanced Optical Technologies (SAOT), Universit\"at Erlangen-N\"urnberg, 91052 Erlangen, Germany}

\author{D.~Bhatti$^{1}$}
\affiliation{$^{1}$Institut f\"{u}r Optik, Information und Photonik, Universit\"{a}t Erlangen-N\"{u}rnberg, 91058 Erlangen, Germany}
\affiliation{$^{2}$Erlangen Graduate School in Advanced Optical Technologies (SAOT), Universit\"at Erlangen-N\"urnberg, 91052 Erlangen, Germany}

\author{J.~von~Zanthier$^{1,2}$}
\affiliation{$^{1}$Institut f\"{u}r Optik, Information und Photonik, Universit\"{a}t Erlangen-N\"{u}rnberg, 91058 Erlangen, Germany}
\affiliation{$^{2}$Erlangen Graduate School in Advanced Optical Technologies (SAOT), Universit\"at Erlangen-N\"urnberg, 91052 Erlangen, Germany}

\author{G.~S.~Agarwal}
\affiliation{Department of Physics, Oklahoma State University, Stillwater, OK 74078, USA}

\date{\today}

\begin{abstract}
Superradiance typically requires preparation of atoms in highly entangled multi-particle states, the so-called Dicke states. In this paper we discuss an alternative route where we prepare such states from initially uncorrelated atoms by a measurement process. By measuring higher order intensity intensity correlations we demonstrate that we can simulate the emission characteristics of Dicke superradiance by starting with atoms in the fully excited state. We describe the essence of the scheme by first investigating two excited atoms. Here we demonstrate how via Hanbury Brown and Twiss type of measurements we can produce Dicke superradiance and subradiance displayed commonly with two atoms in the single excited symmetric and antisymmetric Dicke states, respectively.
We thereafter generalize the scheme to arbitrary numbers of atoms and detectors, and explain in detail the mechanism which leads to this result. The approach shows that Hanbury Brown and Twiss type intensity interference and the phenomenon of Dicke superradiance can be regarded as two sides of the same coin. We also present a compact result for the characteristic functional which generates all order intensity intensity correlations.
\end{abstract}


\maketitle

\section{Introduction}

Modification of spontaneous decay is one of the fundamental topics in quantum physics. In a pioneering paper, Dicke introduced in 1954 the concept of superradiance, i.e., the coherent emission of spontaneous radiation \cite{Dicke(1954)}. The phenomenon is displayed by quantum systems in particular correlated states, the so-called Dicke states, leading to profound modifications of the temporal, directional and spectral emission characteristics of the ensemble compared to that of a single atom \cite{Dicke(1954),Eberly(1971),Manassah(1973),Haroche(1982),AGARWAL(1974),Scully(2006),*Scully(2007),*Scully(2008),*Scully(2009),*Scully(2009)Super,Kaiser(2012),*Kaiser(2013),Nori(2011),*Nori(2012),Nienhuis(1987),Keitel(2015),Cirac(2010),*Cirac(2011),*Cirac(2012),*Cirac(2015),Rohlsberger(2010),*Rohlsberger(2013),Evers(2014),Scully(2015),*Scully(2015)2,Felinto(2014),Kuzmich(2003)}. Even though a tremendous amount of literature - both, theoretical and experimental - has been published since, the physical origin of the phenomenon remained largely  obscure. A deeper understanding has emerged recently in terms of quantum interferences among multiple path ways produced by systems in correlated states \cite{Wiegner(2011)PRA}. Based on this interpretation we investigate in the present paper a new aspect of superradiance, namely that it can be observed also with statistically independent and initially uncorrelated incoherent sources \cite{Oppel(2014)}. 

In our approach the production of atomic correlations and the corresponding superradiant behavior relies on the successive measurement of photons at particular positions in the far field of the sources, 
such that the detection is unable to identify the individual photon source.
In this case, 
the  initially fully excited uncorrelated atomic system cascades down the ladder of symmetric Dicke states each time a photon is recorded. This is another example of measurement induced entanglement among parties which do not  interact with each other and are separated even by macroscopic distances \cite{Cabrillo(1999),*Skornia(2001),*Kimble(2005),*Monroe(2007),*Kimble(2010),*Weinfurter(2012),*Hanson(2013),Thiel(2007)}. 
As discussed below, recording $m$ photons scattered from $N \geq m$ atoms amounts to measuring the $m$-th order photon correlation function. Measuring this function thus allows (a) to produce any desired symmetric Dicke state from statistically independent and initially uncorrelated incoherent atomic sources and (b) to observe the corresponding superradiant emission characteristics of the related Dicke state. 


The setup employed to display the superradiant emission characteristics of initially uncorrelated incoherent atomic sources is similar to the one used in the celebrated Hanbury Brown and Twiss experiment to measure the angular diameter or the separation of stars \cite{HBT(1956)Correlation,HANBURY(1974)}. 
The detailed analysis shows that the two effects derive indeed from the same cause, namely from multi-photon interferences appearing in the $m$-th order photon correlation function. In this way we show that Hanbury Brown and Twiss intensity interference and the phenomenon of Dicke superradiance can be regarded as being two sides of the same coin \cite{Oppel(2014)}. Additionally, in a sense, we provide the great utility of Glauber's program \cite{Glauber(1963)Quantum,*Glauber(1963)Coherent} to extract complete quantum statistical information on radiation fields by means of higher order photon correlation measurements. Last but not least, we show how Dicke superradiance can be simulated by the measurements of higher order intensity correlations on statistically independent and initially uncorrelated incoherent sources. 

The paper is organized as follows. In Sec.~\ref{sec:HBT_2Atoms} we investigate in detail the second order intensity correlation function for two independent atoms in the fully excited state. 
We show that the function displays the same emission characteristics as Dicke super- and subradiance for two atoms in the single excited symmetric and antisymmetric Dicke state, respectively. It is demonstrated that the detection of the first photon can be regarded as a projection of the initially uncorrelated system onto one of the maximally entangled Dicke states, and the detection of the second photon as a probe of the super- and subradiant emission behavior of the corresponding Dicke state.
In Sec.~\ref{sec:GmNAtoms} we generalize this idea and investigate the $m$-th order correlation function for $N$ arbitrary fully excited independent atoms. Considering all possible multi-photon quantum paths leading to a valid $m$-photon detection event we show that if $(m-1)$ detectors are placed at the same position we obtain again a superradiant intensity distribution. In Sec.~\ref{sec:Intensity} we compare the results of Sec.~\ref{sec:GmNAtoms} to the intensity measurement of $N$ atoms being prepared in symmetric Dicke states. We demonstrate that the intensity distribution of a Dicke state with $N-(m-1)$ excited atoms corresponds indeed to the $m$-th order correlation function for the initially fully excited state -- what proves the state projection by photon subtraction. In particular for $m=N$, we prepare effectively the Dicke state with a single excitation and the measurement of the $N$-th order correlation becomes equivalent to the measurement of $G^{(1)}$ for the state with a single excitation.
Sec.~\ref{sec:functional} introduces the characteristic functional, which can be used for calculating higher order intensity correlation functions in a compact way. For $(m-1)$ detectors at the same position identical results as in Sec.~\ref{sec:GmNAtoms} are obtained. In Sec.~\ref{sec:conclusion} we finally conclude.
Note that in a subsequent paper we will apply the ideas of this paper to classical sources and discuss in detail the aspects of superradiant emission from such kind of emitters.

\section{Hanbury Brown and Twiss Effect and Dicke Super- and Subradiance with Two Radiating Atoms}
\label{sec:HBT_2Atoms}

\begin{figure}[b]
\centering
\includegraphics[width=0.4\textwidth]{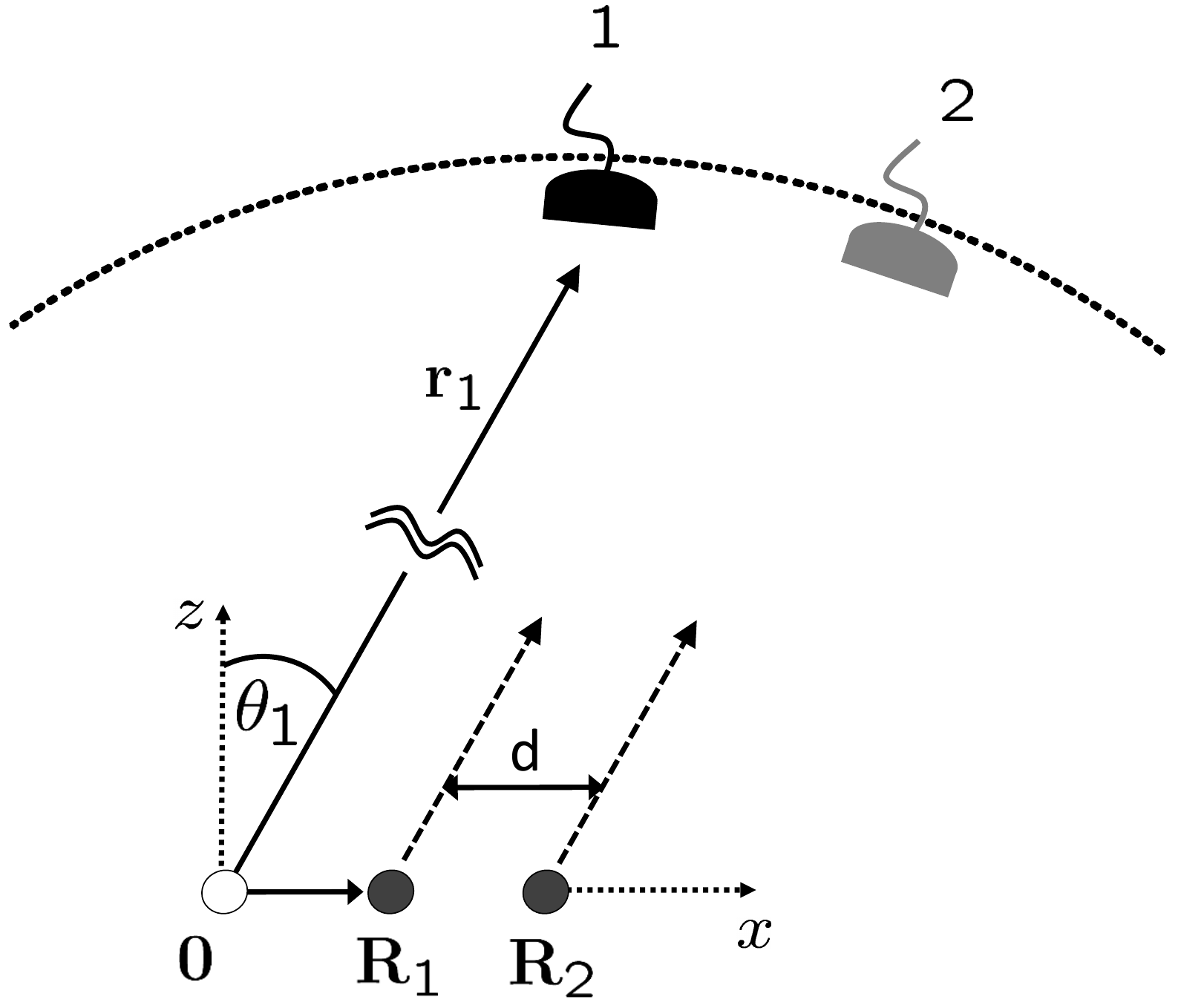}
\caption{Considered setup: Two identical two-level atoms, separated by a distance $d \gg \lambda$, are placed at positions $\vec{R}_{l}$, $l = 1, 2$; the light scattered by the sources is measured by two detectors, located at positions $\vec{r}_j$, $j = 1, 2$ in the far field.}
\label{fig:detection_scheme}
\end{figure}

The classical Hanbury Brown and Twiss effect was measured with classical thermal sources. The measured quantity was the spatial intensity intensity correlation $\left< I(\vec{r}_{1})I(\vec{r}_{2}) \right>$ as recorded by two detectors located at $\vec{r}_{1}$ and $\vec{r}_{2}$. Now we consider the Hanbury Brown and Twiss measurement with two independent fully excited  atoms at ${\vec R}_{1}$ and ${\vec R}_{2}$, separated by distance $d$ much larger than the wavelength $\lambda$ of the emitted photons, such that the dipole-dipole coupling between the atoms can be neglected. With reference to Fig.~\ref{fig:detection_scheme} we consider the two detectors positioned in the far field in a circle around the sources. For simplicity we suppose that the atoms and the detectors are in one plane and that the atomic dipole moments are oriented perpendicular to this plane. The measured quantity is
\begin{equation}
	G^{(2)}(\vec{r}_{1},\vec{r}_{2})= \left< \mathcal{E}^{(-)}(\vec{r}_{1})\! \cdot\!  \mathcal{E}^{(-)}(\vec{r}_{2})\! \cdot\! \mathcal{E}^{(+)}(\vec{r}_{2})\! \cdot\! \mathcal{E}^{(+)}(\vec{r}_{1}) \right> \ ,
\end{equation}
where $\mathcal{E}^{(+)}(\vec{r}_{1})$ and $\mathcal{E}^{(-)}(\vec{r}_{1})$ are the positive and negative frequency parts of the electric field operator, respectively.

Each two-level atom is represented by the spin $1/2$ operators $\vec{s}^{\, (l)}$ ($l=1,2$). The field at the detector position can be related to the atomic operators via the well known relation \cite{AGARWAL(2013)}
\begin{equation}
\begin{aligned}
	\vec{\mathcal{E}}^{(+)}(\vec{r}_{1},t)&\sim   \ \mathcal{E}_{0}^{(+)}(\vec{r}_{1},t)  \\ 
	- \frac{\omega^{2}}{c^{2}} & \frac{e^{i(\frac{\omega}{c}r_{1} - \omega t)} }{r_{1}} \sum_{l=1,2} e^{-i \frac{\omega}{c}\vec{n}_{1}\cdot \vec{R}_{l}}\, s_{-}^{ (l)} (\vec{n}_{1}\times(\vec{n}_{1}\times\vec{p})) \ ,
\label{eq:field}
\end{aligned}
\end{equation}
where $\vec{p}$ is the transition dipole moment, $\vec{n}_{1}=\vec{r}_{1}/|\vec{r}_{1}|$ is the direction of observation, $\mathcal{E}_{0}^{(+)}$ is the free field operator and $s_{-}^{(l)}=s_{x}^{(l)}-is_{y}^{(l)}$ ($s_{-}^{(l)}=s_{x}^{(l)}+i s_{y}^{(l)}$) is the atomic lowering (raising) operator for atom $l$.

With the initial state $\ket{\Phi}=\ket{e,e}$ and using Eq.~(\ref{eq:field}), the intensity intensity correlation function can be calculated to be
\begin{equation}
	G^{(2)}(\vec{r}_{1},\vec{r}_{2})= \frac{\omega^{8}}{c^{8}r_{1}^{2}r_{2}^{2}} |(\vec{n}_{1}\times\vec{p})|^{2} |(\vec{n}_{2}\times\vec{p})|^{2} \bar{G}^{(2)}(\vec{r}_{1},\vec{r}_{2})   \ ,
\end{equation}
where
\begin{equation}
	\bar{G}^{(2)}(\vec{r}_{1},\vec{r}_{2}) = 2 [1+ \cos\left(\frac{\omega}{c} (\vec{n}_{1}-\vec{n}_{2})\cdot(\vec{R}_{1}-\vec{R}_{2})\right)]  \ ,
\label{eq:Gbar}
\end{equation}
with $\vec{n}_{j}=\vec{r}_{j}/|\vec{r}_{j}|$ ($j=1,2$) denoting the two unit vectors pointing towards the two detectors.
In what follows we work with $\bar{G}^{(2)}$ which effectively is defined by
\begin{equation}
	\bar{G}^{(2)}(\vec{r}_{1},\vec{r}_{2}) = \left< E^{(-)}(\vec{r}_{1})E^{(-)}(\vec{r}_{2})E^{(+)}(\vec{r}_{2})E^{(+)}(\vec{r}_{1}) \right> \ , 
\label{eq:Gbar2}
\end{equation}
with
\begin{equation}
\left[ E^{(-)}(\vec{r}) \right]^{\dagger}	= E^{(+)}(\vec{r}) =\sum_{l} e^{-i\frac{\omega}{c}\vec{n}\cdot \vec{R}_{l}} s_{-}^{(l)} \, .
\label{eq:Eplus}
\end{equation}
For brevity we will drop the bar from $\bar{G}^{(2)}$.
Note the presence of fringes in the intensity intensity correlations (Eq.~(\ref{eq:Gbar})) as the detector positions are varied. 

The normalized form of the intensity intensity correlation reads \cite{Skornia(2002)}
\begin{equation}
\begin{aligned}
	g^{(2)}(\vec{r}_{1},\vec{r}_{2})&= \frac{G^{(2)}(\vec{r}_{1},\vec{r}_{2})}{G^{(1)}(\vec{r}_{1})G^{(1)}(\vec{r}_{2})} \\
		&= \frac{1}{2} [1 + \cos\left(\frac{\omega}{c} (\vec{n}_{1}-\vec{n}_{2})\cdot(\vec{R}_{1}-\vec{R}_{2})\right)] \ ,
\label{eq:g2}
\end{aligned}
\end{equation}
with
\begin{equation}		
		G^{(1)}(\vec{r}) = \left< E^{(-)}(\vec{r}) E^{(+)}(\vec{r}) \right> = 2 \ .
\label{eq:G1}
\end{equation}
Note that there are some similarities to the Hanbury Brown and Twiss result for thermal sources \cite{MANDEL(1995)}
\begin{equation}
\begin{aligned}
	g^{(2)}_{thermal}(\vec{r}_{1},\vec{r}_{2})= 1 + |\gamma(\vec{r}_{1}-\vec{r}_{2})|^{2} \ ,
\label{eq:g2thermal}
\end{aligned}
\end{equation}
where $\gamma(\vec{r}_{1}-\vec{r}_{2})$ is the complex degree of first order coherence between the two thermal sources. However, unlike Eq.~(\ref{eq:g2thermal}), we could have 
\begin{equation}
	g^{(2)} \rightarrow 0 \ \ \text{for single atom sources,}
\end{equation}
i.e., the probability of detecting one photon at each of the detectors $D_{1}$ and $D_{2}$ can be zero in case that
\begin{equation}
	\frac{\omega}{c} (\vec{n}_{1}-\vec{n}_{2})\cdot(\vec{R}_{1}-\vec{R}_{2}) = \pi \ .
\end{equation}
This is the well known Hong-Ou-Mandel effect \cite{HOM(1987)} with two identical single photons - with the single photons being emitted by the identically excited atoms. We have thus established a connection between the Hanbury Brown and Twiss effect and the Hong-Ou-Mandel effect for two identical single photon sources \footnote{A detailed discussion of this aspect for arbitrary numbers of single photon sources will be presented in \cite{Maehrlein(2015)}}.

Dicke superradiance and subradiance require that the two atoms are prepared in an entangled state. Let us consider the initially prepared state
\begin{equation}
	\ket{\Psi}= \frac{1}{\sqrt{2}} \left( \ket{e,g} + e^{i \delta}\ket{g,e} \right) \ ,
\label{eq:Psi}
\end{equation}
where $\delta$ is a parameter defining for $\delta=0$ ($\pi$) a symmetric (antisymmetric) state.

From Eq.~(\ref{eq:Psi}), the atomic expectation values are given by
\begin{equation}
	\left< s_{+}^{(1)}s_{-}^{(2)} \right> = \frac{1}{2}e^{i \delta} \ , \ \ \left< s_{+}^{(j)}s_{-}^{(j)} \right> = \frac{1}{2} \ .
\label{eq:s+s-half}
\end{equation}
Using Eq.~(\ref{eq:s+s-half}) the intensity of the emission becomes
\begin{equation}
\begin{aligned}
I &=\left< E^{(-)}(\vec{r}_1) E^{(+)}(\vec{r}_1) \right > \\ 
&= \left< s_{+}^{(1)}s_{-}^{(1)} \right> + \left< s_{+}^{(2)}s_{-}^{(2)} \right> + 2 \Re{\left< s_{+}^{(1)}s_{-}^{(2)} \right> e^{i \frac{\omega}{c}\vec{n}_1 \cdot (\vec{R}_{1}-\vec{R}_{2})}} \\
&= 1 + \cos[\delta+ \frac{\omega}{c}\vec{n}_1 \cdot (\vec{R}_{1}-\vec{R}_{2})] \ .
\label{eq:I0}
\end{aligned}
\end{equation}
The intensity distribution exhibits a fringe pattern as the detector position is scanned. 
In particular, for a detection perpendicular to the line joining the atoms, we find $I = 2$ for $\delta=0$ and $I = 0$ for $\delta=\pi$ (a similar result is obtained if the two atoms are confined to a region much smaller than the wavelength $\lambda$, i.e., $|\frac{\omega}{c}\vec{n}_{1}\cdot (\vec{R}_{1}-\vec{R}_{2})| \ll 1$).
It should be borne in mind that there is only a single excitation in the system. Thus for perpendicular detection (or for two atoms separated by $d \ll \lambda$), we obtain Dicke superradiance and subradiance for $\delta=0$ and $\delta=\pi$, respectively. 
For these values, the states of Eq.~(\ref{eq:Psi}) are the Dicke states $\ket{j,m}$ with $j=1, m=0$ for $\delta=0$ and $j=0, m=0$ for $\delta=\pi$.

If we compare Eq.~(\ref{eq:g2}) with Eq.~(\ref{eq:I0}) we see that the fringe patterns in Eq.~(\ref{eq:I0}) and Eq.~(\ref{eq:g2}) are identical if we take
\begin{equation}
	\delta = - \frac{\omega}{c} \vec{n}_{2} \cdot (\vec{R}_{1}-\vec{R}_{2}) \ .
\label{eq:delta}
\end{equation}
Clearly there must then be a deep reason for this to hold. It implies that a measurement of $G^{(1)}(\vec{r}_{1})$ with the initial state given by Eq.~(\ref{eq:Psi}), along with Eq.~(\ref{eq:delta}), is the same as a measurement of $G^{(2)}(\vec{r}_{2},\vec{r}_{1})$ for the initial state $\ket{e,e}$. This clearly has an important consequence as it shows that the physics of the entangled state Eq.~(\ref{eq:Psi}) can be explored via a measurement of $G^{(2)}$ on the state $\ket{e,e}$. Thus the measurement of $G^{(2)}$ bypasses the need for the preparation of the entangled state Eq.~(\ref{eq:Psi}) as the superradiant and subradiant characteristics of Eq~(\ref{eq:Psi}) can be studied via the $G^{(2)}$ measurement on the state $\ket{e,e}$.

We next present a mathematical reasoning for this finding. Let us write $G^{(2)}(\vec{r}_{2},\vec{r}_{1})$ for the initial state $\ket{\Phi}=\ket{e,e}$ in the form
\begin{equation}
\begin{aligned}
	G_{\Phi}^{(2)}(\vec{r}_{2},\vec{r}_{1})&= \bra{\Phi} E^{(-)}(\vec{r}_{2})E^{(-)}(\vec{r}_{1})E^{(+)}(\vec{r}_{1})E^{(+)}(\vec{r}_{2}) \ket{\Phi} \\
		&= \left( \text{Tr}\left[\tilde{\rho}  E^{(-)}(\vec{r}_{1})E^{(+)}(\vec{r}_{1})\right] \right) G_{\Phi}^{(1)}(\vec{r}_{2}) \ .
\label{eq:G2Phi}
\end{aligned}
\end{equation}
where we introduce
\begin{equation}
\begin{aligned}
	\tilde{\rho}= \frac{ E^{(+)}(\vec{r}_{2}) \ket{\Phi}\bra{\Phi} E^{(-)}(\vec{r}_{2}) }{G_{\Phi}^{(1)}(\vec{r}_{2})} \ \ , \ \text{Tr}[\tilde{\rho}] = 1 \ .
\label{eq:rho}
\end{aligned}
\end{equation}
and
\begin{equation}
	G_{\Phi}^{(1)}(\vec{r}_{2})= \bra{\Phi} E^{(-)}(\vec{r}_{2})E^{(+)}(\vec{r}_{2}) \ket{\Phi} \ .
\label{eq:G1Phi}
\end{equation}
Rewriting Eq.~(\ref{eq:G2Phi}) in the form
\begin{equation}
	G_{\Phi}^{(2)}(\vec{r}_{2},\vec{r}_{1})= G_{\tilde{\rho}}^{(1)}(\vec{r}_{1})G_{\Phi}^{(1)}(\vec{r}_{2}) 
\label{eq:G2Phi2}
\end{equation}
leads to the following important result: A measurement $G_{\Phi}^{(2)}(\vec{r}_{2},\vec{r}_{1})/G_{\Phi}^{(1)}(\vec{r}_{2})$ on the state $\ket{\Phi}=\ket{e,e}$ is equivalent to a measurement of $G_{\tilde{\rho}}^{(1)}(\vec{r}_{1})$ on the state $\tilde{\rho}$ which is determined by the location of the detector $D_{2}$. Note that according to Eq.~(\ref{eq:G2Phi2}), we can identify
\begin{equation}
G_{\tilde{\rho}}^{(1)}(\vec{r}_{1}) = G_{\Phi,c}^{(2)} (\vec{r}_{2},\vec{r}_{1}) \, ,
\label{eq:G2conditional}
\end{equation}
as the conditional intensity intensity correlation, i.e., the joint probability of detecting one photon at $\vec{r}_{1}$ conditioned on the detection of a photon at $\vec{r}_{2}$. 
The state $\tilde{\rho}$ can be found using Eq.~(\ref{eq:Eplus}) in Eq.~(\ref{eq:rho}) with the result
\begin{equation}
\begin{aligned}
	\tilde{\rho} &= \ket{\chi}\bra{\chi} \  , \\
	\ket{\chi} &= \frac{1}{\sqrt{2}} \left( \ket{g,e} e^{-i\frac{\omega}{c} \vec{n}_{2}\cdot \vec{R}_{1}} + \ket{e,g} e^{-i\frac{\omega}{c} \vec{n}_{2}\cdot \vec{R}_{2}} \right) \ .
\label{eq:rho2}
\end{aligned}
\end{equation}
We have thus proved exactly the isomorphism between $G_{\Phi}^{(2)}(\vec{r}_{2},\vec{r}_{1})$ and $G_{\tilde{\rho}}^{(1)}(\vec{r}_{1})$. The explicit form of the normalized conditional intensity intensity correlation $G_{\Phi,c}^{(2)} (\vec{r}_{2},\vec{r}_{1})$ reads (cf. Eq.~(\ref{eq:G2conditional}))
\begin{equation}
\begin{aligned}
	G_{\Phi,c}^{(2)} (\vec{r}_{2},\vec{r}_{1}) =\! \left[1 + \cos\left( \frac{\omega}{c} (\vec{n}_{1}-\vec{n}_{2})\cdot(\vec{R}_{1}-\vec{R}_{2}) \right)\right] .
\end{aligned}
\label{eq:normg2}
\end{equation}
Clearly, according to Eq.~(\ref{eq:normg2}), we have
\begin{equation}
\left(G_{\Phi,c}^{(2)} \right)_{\text{max}}=2 \ \ , \ \ \left(G_{\Phi,c}^{(2)} \right)_{\text{min}}=0 \ , 
\end{equation}
which are clear signatures of superradiance and subradiance from the two atom system. A great advantage of this strategy is that we need not to prepare the system in an entangled state.

The special case discussed above is applicable more generally, and in particular is not restricted to single photon sources. To demonstrate that let us think of the radiation field in the state $\rho$. Then $G_{\rho}^{(2)}(\vec{r}_{2},\vec{r}_{1})$ can be written as
\begin{equation}
\begin{aligned}
	G_{\rho}^{(2)}(\vec{r}_{2},\vec{r}_{1}) &= \text{Tr}[\rho \, E^{(-)}(\vec{r}_{2})E^{(-)}(\vec{r}_{1})E^{(+)}(\vec{r}_{1})E^{(+)}(\vec{r}_{2})] \\
	& = G_{\tilde{\rho}}^{(1)}(\vec{r}_{1}) G_{\rho}^{(1)}(\vec{r}_{2}) = G_{\rho,c}^{(2)}(\vec{r}_{2}, \vec{r}_{1}) G_{\rho}^{(1)}(\vec{r}_{2}) \ , \\
\end{aligned}
\end{equation}
with
\begin{equation}
	G_{\rho,c}^{(2)}(\vec{r}_{2}, \vec{r}_{1}) = G_{\tilde{\rho}}^{(1)}(\vec{r}_{1}) \ ,
\end{equation}
where
\begin{equation}
\tilde{\rho} = \frac{E^{(+)}(\vec{r}_{2}) \rho E^{(-)}(\vec{r}_{2})}{\text{Tr}[ \rho E^{(-)}(\vec{r}_{2})E^{(+)}(\vec{r}_{2}) ]} \ .
\end{equation}
Thus the conditional intensity intensity correlation $	G_{\rho,c}^{(2)}(\vec{r}_{2}, \vec{r}_{1})$ is again the same as $G_{\tilde{\rho}}^{(1)}(\vec{r}_{1})$ of the projected state $\tilde{\rho}$. Note that $E^{(+)}(\vec{r}_{2})$ is the annihilation operator and hence $\tilde{\rho}$ is obtained by subtracting a photon from the state $\rho.$ Thus $G_{\rho}^{(2)}(\vec{r}_{2},\vec{r}_{1})$ can be thought of providing information on the photon subtracted state $\tilde{\rho}$. It should be born in mind that the projected state $\tilde{\rho}$ depends on the location at which the photon is detected.

\section{\texorpdfstring{$m$}{}-th order correlation function for \texorpdfstring{$N$}{} single photon emitters}
\label{sec:GmNAtoms}

\begin{figure}[t]
\centering
\includegraphics[width=0.4\textwidth]{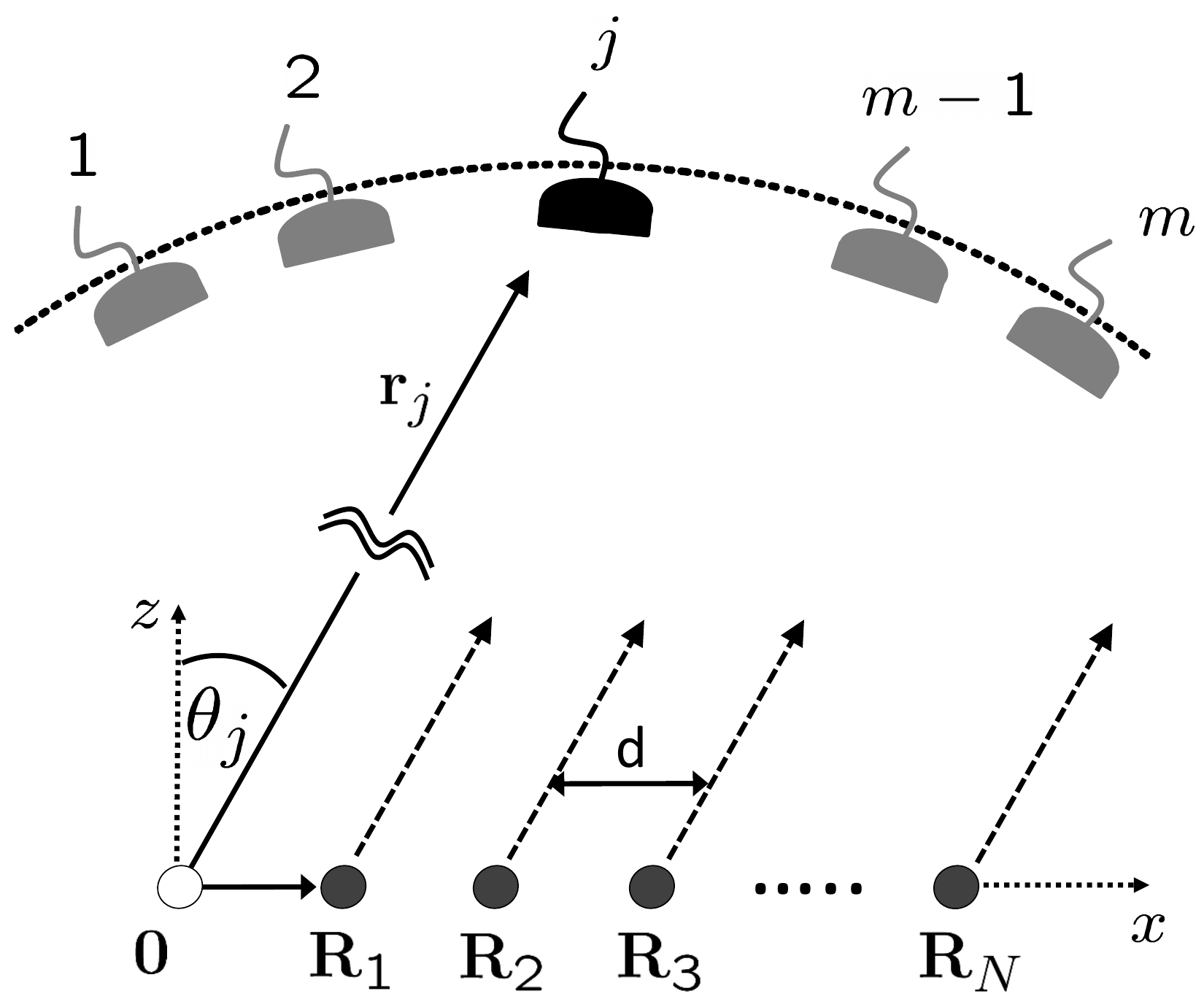}
\caption{Considered setup: $N$ identical two-level atoms, separated by a distance $d \gg \lambda$, are placed along a chain at positions ${\vec R}_{l}$, $l = 1, \ldots , N$; the light scattered by the sources is measured by $m$ detectors, located at positions ${\vec r}_j$, $j = 1, \ldots , m$, in the far field.}
\label{fig:scheme_N}
\end{figure}

In this section we generalize the ideas of Sec.~\ref{sec:HBT_2Atoms} and derive the $m$-th order correlation function for $N$ identical single photon emitters (SPE), e.g., two-level atoms with upper state $\ket{e_{l}}$ and ground state $\ket{g_{l}}$, assuming that $(m-1)$ spontaneously emitted photons are detected at $\vec{r}_1$ and the $m$-th photon at $\vec{r}_2$. The $m$ detectors are supposed to be located at positions ${\vec r}_j$, $j =  1, \ldots ,m$, in the far field in a circle around the sources and the emitters at positions ${\vec R}_{l}$, $l = 1, \ldots, N$, along a linear chain with equal spacing $d \gg \lambda$ such that the dipole-dipole coupling between them can be neglected (see~Fig.~\ref{fig:scheme_N}). 
For simplicity we suppose again that the emitters and the detectors are in one plane and that the atomic dipole moments of the transition $\ket{e_{l}} \rightarrow \ket{g_{l}}$ are oriented perpendicular to this plane. Note that we assume a single photon counting regime, where all $m$ photons have to be detected for a valid measurement. That is via postselection we keep only those events where all $m$ detectors click, all other events are dropped. This is not a handicap as what matters for our theoretical analysis is the conditional measurement of the $m$-th photon.

For arbitrary detector positions the $m$-th order correlation function is defined as 
\staf
\label{Eq1n}
G^{(m)} (\vec{r}_1, \ldots , \vec{r}_m)  \equiv \average{:\prod_{j=1}^m E^{(-)}(\fr_j) E^{(+)}(\fr_j):} \, ,
\stof
where $\average{: \ldots :}$ denotes the (normally ordered) quantum mechanical expectation value. Due to the far field condition, i.e., the inability to identify the particular photon sources, the electric field operator at ${\vec r}_j$ is given by \cite{Thiel(2007)}
\staf
\label{Efield}
\left[ E^{(-)}(\vec{r}_j) \right]^\dagger = E^{(+)}(\vec{r}_j) \sim \sum_{l=1}^{N} e^{-i\,\varphi_{lj}} \;\hat{s}^{-}_l \, .
\stof
Here, $\hat{s}^{-}_l = |g_l\rangle\langle e_l|$ is the atomic lowering operator and 
\staf
\label{varphi_lj}
\varphi_{lj}  = k \, \frac{\vec{r}_j\cdot\vec{R}_l}{r_j} = l\,kd\,\sin\theta_j
\stof
denotes the optical phase accumulated by a photon emitted at $\vec{R}_l$ and detected at ${\vec r}_j$ relative to a photon emitted at the origin (cf.~Fig.~\ref{fig:scheme_N}).
Note that for simplicity we define as in the previous section the field and hence all correlation functions of $m$-th order dimensionless; the actual values can be obtained by multiplying $G^{(m)}$ with $m$ times the intensity of a single source.

\begin{figure}[b]
  \centering
	\includegraphics[width=0.48\textwidth]{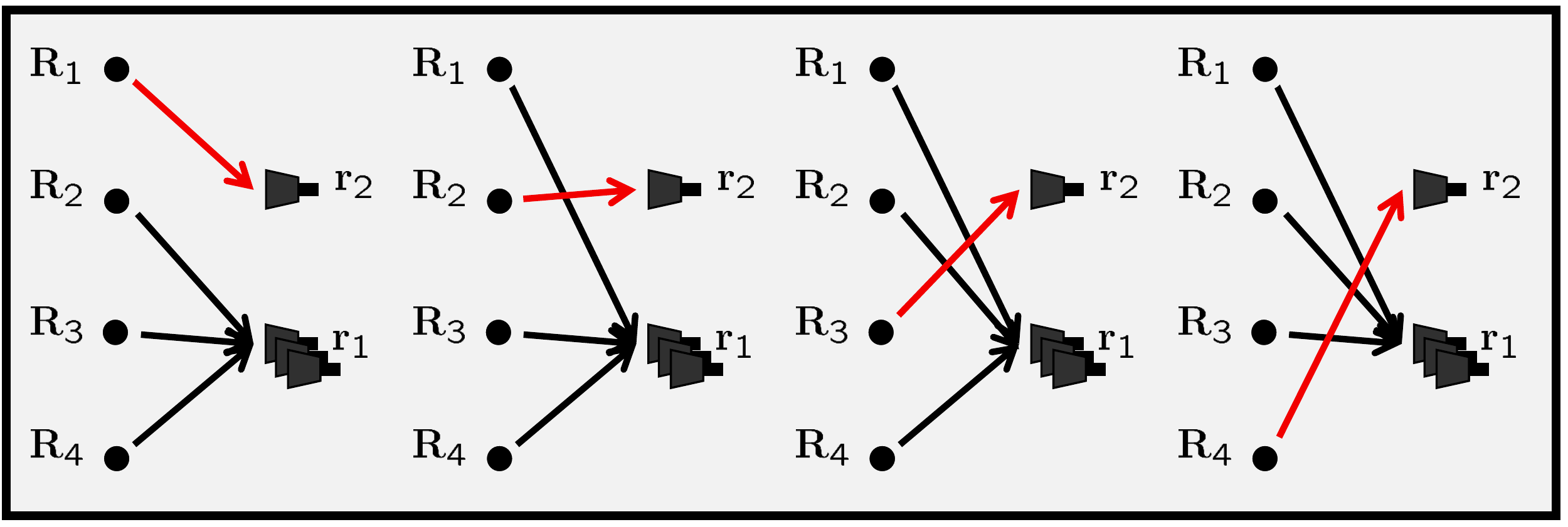}
  \caption{(Color online) $m$-photon quantum paths of $N$ statistically independent SPE where $(m-1)$ photons are detected at $\vec{r}_1$ and one photon at $\vec{r}_2$ (see~Eq.~(\ref{GMGENSb2})). The interference signal $ G^{(m)}_{\Phi_{N}} (\vec{r}_1, ...,\vec{r}_1,\vec{r}_2)$ can be reduced to a superposition of $m$ indistinguishable yet different $m$-photon quantum paths. In the figure the explicit case of $N=m=4$ sources and detectors is shown.}
  \label{fig:paths_gm_4}
\end{figure}

Starting with all atoms in the excited state, i.e., in the uncorrelated state $\ket{\Phi_{N}}  \equiv \prod_{l = 1}^{N} \ket{e_l}$,
we find from Eq.~(\ref{Eq1n}) for the $m$-th order correlation function
\begin{equation}
\begin{aligned}
\label{GMGENSa}
& \mathrm{G}^{(m)}_{\Phi_{N}}(\vec{r}_1,..., \vec{r}_m)  \\
& \sim \Big\|\sum_{\sigma_{1}=1}^{N} \sum_{\substack{\sigma_{2}=1 \\ \sigma_1 \neq\sigma_2}}^{N} \hdots \sum_{\substack{\sigma_m = 1 \\ \sigma_i \neq \sigma_{m};\, i< m}}^{N} \prod_{j = 1}^{m} e^{-i\,\varphi_{\sigma_j j}} \ket{g_{\sigma_j}}\Big\|^2  \\
& = \sum_{\sigma_{1}=1}^{N} \sum_{ \sigma_{2}=\sigma_1+1 }^{N} \hdots \sum_{\sigma_{m}=\sigma_{m-1}+1}^{N} \Big|\sum_{\substack{\sigma_1,...,\sigma_m \\\in \,{\cal S}_m}}\prod_{j = 1}^{m} e^{-i\,\varphi_{\sigma_j j}}\Big|^2\, .
\end{aligned}
\end{equation}
Here, $\|\ket{\psi}\|^2 = \langle \psi | \psi \rangle$ defines the norm of the state vector $\ket{\psi}$, $|...|$ abbreviates absolute values, and the expression $\sum_{\substack{\sigma_1,...,\sigma_m \\\in \,{\cal S}_m}}$ denotes the sum over the symmetric group ${\cal S}_m$ with elements $\sigma_1,...,\sigma_m$. 
In Eq.~(\ref{GMGENSa}) the products $\prod_{j = 1}^{m} e^{-i\,\varphi_{\sigma_j j}}$ represent $m$-photon quantum paths with phases $\sum_{j=1}^{m} \varphi_{\sigma_j j}$, accumulated by $m$ photons emitted from $m$ sources at $\vec{R}_{\sigma_j}$ and recorded by $m$ detectors at ${\vec r}_j$. Since the particular source of a recorded photon is unknown we have to sum over all possible combinations of $m$-photon quantum paths, what is expressed by the sum 
$\sum_{\sigma_{1}=1}^{N} \sum_{\sigma_{2}=1}^{N} \hdots \sum_{\sigma_{m}=1}^{N}$ in the second line of Eq. (\ref{GMGENSa}). Hereby, the additional condition
$\sigma_{i}\neq \sigma_{j}$ ($i\neq j$) ensures that each source emits at most one photon.
Considering that several combinations of $m$-photon quantum paths lead to the \textit{same} final atomic state and thus have to be added coherently, we end up with the modulus square in the third line of Eq.~(\ref{GMGENSa}). 
Hereby, for the $\binom{N}{m}$ \textit{different} final atomic states, the corresponding transition probabilities $|...|^2$ have to be summed incoherently, what results in the first combination of sums 
$\sum_{\sigma_{1}=1}^{N} \sum_{ \sigma_{2}=\sigma_1+1 }^{N} \hdots \sum_{\sigma_{m}=\sigma_{m-1}+1}^{N}$ of the third line of Eq.~(\ref{GMGENSa}). 

We next consider that $(m-1)$ detectors are placed at the same position ${\vec r}_1$ and the last detector at ${\vec r}_2$ (see Fig.~\ref{fig:paths_gm_4}). 
Under these conditions we can make use of the fact that out of the $m!$ $m$-photon quantum paths, denoted by $\sum_{\substack{\sigma_1,...,\sigma_m \\\in \,{\cal S}_m}}\prod_{j = 1}^{m} e^{-i\,\varphi_{\sigma_j j}}$ in Eq.~(\ref{GMGENSa}), $(m-1)!$ quantum paths are equal since $(m-1)$ detectors are placed at the same position. In this case Eq.~(\ref{GMGENSa}) takes the form 
\staffeld
\label{GMGENSb2}
&{G}^{(m)}_{\Phi_{N}}(\vec{r}_1,..., \vec{r}_1, \vec{r}_2)  \nonumber\\
& = \sum_{\substack{\sigma_1, ..., \sigma_m = 1 \\ \sigma_1 < ... <  \sigma_m}}^{N} |(m-1)! \sum_{j=1}^m e^{-i\,(\varphi_{\sigma_j 2} - \varphi_{\sigma_j 1})}|^2 \nonumber\\
& = ((m-1)!)^{2} \sum_{\substack{\sigma_1, ..., \sigma_m = 1 \\ \sigma_1 < ... <  \sigma_m}}^{N}|\sum_{j=1}^m e^{-i\,(\varphi_{\sigma_j 2} - \varphi_{\sigma_j 1})}|^2\nonumber\\
&= ((m-1)!)^{2} \binom{N}{m} \left( \frac{m\,(N-m)}{N-1} + \frac{m\,(m-1)}{N\,(N-1)} \frac{\sin^2(N\frac{\varphi_{11}-\varphi_{12}}{2})}{\sin^2(\frac{\varphi_{11}-\varphi_{12}}{2})}\right)\,, & \nonumber\\
\stoffeld
with $\sum_{\substack{\sigma_1, ..., \sigma_m = 1 \\ \sigma_1 < ... <  \sigma_m}}^{N}=\sum_{\sigma_{1}=1}^{N} \sum_{ \sigma_{2}=\sigma_1+1 }^{N} \hdots \sum_{\sigma_{m}=\sigma_{m-1}+1}^{N}$. Finally we obtain for the $m$-th order correlation function \cite{Oppel(2014)}
\staffeld
\label{GMGENSbb}
& {G}^{(m)}_{\Phi_{N}}(\vec{r}_1,..., \vec{r}_1, \vec{r}_2)  \nonumber\\
&= \frac{N!(m-1)!}{(N-m)!}\,\left( \frac{N-m}{N-1} + \frac{m-1}{N(N-1)}\frac{\sin^2(N\frac{\varphi_{11}-\varphi_{12}}{2})}{\sin^2(\frac{\varphi_{11}-\varphi_{12}}{2})}\right)\,.
\stoffeld

According to Eq.~(\ref{GMGENSbb}), the $m$-th order correlation function  as a function of $\vec{r}_{2}$ displays maxima and minima with a visibility of 
\staf
\label{visibilitySPE}
{\cal V}_{SPE} = \frac{m-1}{m+1-\frac{2m}{N}} \; ,
\stof
and the central maximum having an angular width of 
\begin{equation}
\label{peakwidth}
\delta\theta_2 \approx \frac{2\pi}{N\,k\,d}\ , 
\end{equation}
where we have used the relation
\begin{equation}
\delta\varphi_{12}= kd\, \cos\theta_{2} \, \delta\theta_{2} \cong kd\, \delta\theta_{2} \ .
\end{equation}
For growing numbers of emitters an increased focusing of the $m$-th photon in the direction of $\vec{r}_{2}=\vec{r}_1$  is thus observed. 

Note that for $m=1$, the visibility ${\cal V}_{SPE}$ vanishes what illustrates the fact that there is no preferred direction of emission of the first photon, i.e., the atoms emit the light incoherently. For $m=1$ all the atoms are in the excited state and it is known that such a state for short times does not show superradiant behavior. However, for $m > 1$, a superradiant emission and collective decay of the $m$-th photon sets in if it is preconditioned by the measurement of (m-1) photons along a particular direction. For $m = N$ a maximum visibility of ${\cal V}_{SPE} = 100 \%$ is obtained. Note further that in the latter case the height of the central maximum grows as $(N!)^{2}$. On the other hand, we find that
\begin{equation}
\begin{aligned}
	\frac{1}{2\pi} \int &G_{\Phi_{N}}^{(m)}(\vec{r}_{1}, \hdots , \vec{r}_{1}, \varphi_{12})\, \text{d}\varphi_{12} \\
	&= \left< G_{\Phi_{N}}^{(m)} \right>_{\varphi_{12}} = ((m-1)!)^{2} \binom{N}{m-1} (N-m+1)\ .
\end{aligned}
\end{equation}
Hence the normalized $m$-th order correlation function $G_{\Phi_{N}}^{(m)}/\left< G_{\Phi_{N}}^{(m)} \right>_{\varphi_{12}}$ has a central maximum which is proportional to $N$ for $m=N$. The scaling of the maximum as $N$ and the angular width as $1/N$ are the typical features of superradiance from a Dicke state with \textit{one excitation}. This then suggest a close connection between the $m$-th order correlations of a fully excited state with the superradiant character of the suitably prepared states with $(m-1)$ deexcitations. In the next section we demonstrate this relationship in detail, following the procedure we used in Sect.~II for two fully excited atoms. Note that temporal aspects of Dicke superradiance are not addressed in our analysis as they require a separate study where one has to solve the superradiance master equation numerically \cite{AGARWAL(2013)15}. 

\section{Superradiance and \texorpdfstring{$m$}{}-th order intensity correlations}
\label{sec:Intensity}

In order to see the connection between superradiance and the $m$-th order intensity correlation function, we write $G_{\Phi_{N}}^{(m)}(\vec{r}_{1}, \hdots , \vec{r}_{1},\vec{r}_{2})$ as
\begin{equation}
\begin{aligned}
&G_{\Phi_{N}}^{(m)}(\vec{r}_{1}, \hdots , \vec{r}_{1},\vec{r}_{2}) \\
&= \text{Tr}[\tilde{\rho}^{(m-1)}E^{(-)}(\vec{r}_{2})E^{(+)}(\vec{r}_{2})] G_{\Phi_{N}}^{(m-1)}(\vec{r}_{1}, \hdots , \vec{r}_{1})  \ ,
\label{eq:GSdef}
\end{aligned}
\end{equation}
where we defined
\begin{equation}
\begin{aligned}
\tilde{\rho}^{(m-1)} &= \frac{\left(E^{(+)}(\vec{r}_{1})\right)^{m-1} \ket{\Phi_{N}}\bra{\Phi_{N}} \left(E^{(-)}(\vec{r}_{1})\right)^{m-1} }{G_{\Phi_{N}}^{(m-1)}(\vec{r}_{1},\hdots ,\vec{r}_{1})} \\
&= \ket{\Psi_{m-1}} \bra{\Psi_{m-1}} \ ,
\label{eq:rhom-1}
\end{aligned}
\end{equation}
i.e.,
\begin{equation}
\begin{aligned}
G_{\Phi_{N}}^{(m)}&(\vec{r}_{1},\hdots ,\vec{r}_{1},\vec{r}_{2}) \\
&= G_{\Psi_{m-1}}^{(1)}(\vec{r}_{2}) \, G_{\Phi_{N}}^{(m-1)}(\vec{r}_{1}, \hdots , \vec{r}_{1}) \ .
\end{aligned}
\end{equation}
This shows that the variation of $G_{\Phi_{N}}^{(m)}(\vec{r}_{1},\hdots ,\vec{r}_{1},\vec{r}_{2})$ with respect to $\vec{r}_{2}$, i.e., $\theta_{2}$, is a function of the state $\ket{\Psi_{m-1}}$. For simplicity we can set $\theta_{1}=0$. In this case we have
\begin{equation}
\begin{aligned}
E^{(+)}(\vec{r}_{1}) = \sum_{l} s_{-}^{(l)} = J_{-} \ ,
\label{eq:JDef}
\end{aligned}
\end{equation}
where $J_{-}$ is the collective lowering operator as introduced by Dicke \cite{Dicke(1954)}. 
Using Eq.~(\ref{eq:JDef}) in Eq.~(\ref{eq:rhom-1}) we obtain
\begin{equation}
\begin{aligned}
\ket{\Psi_{m-1}} & = \frac{(J_{-})^{m-1}\ket{\Phi_{N}}}{\sqrt{\bra{\Phi_{N}}(J_{+})^{m-1}(J_{-})^{m-1}\ket{\Phi_{N}}}} \\
& = \ket{\frac{N}{2}, \frac{N}{2}-(m-1)} \ ,
\label{eq:Psim-1}
\end{aligned}
\end{equation}
where $\ket{\frac{N}{2},M}$ is the Dicke state with cooperation number $N/2$ and $M$ the projection of the angular momentum on the $z$-axis, i.e., $J_{z}\ket{\frac{N}{2},M}=(\frac{1}{2}[J_{+},J_{-}])\ket{\frac{N}{2},M}=M\ket{\frac{N}{2},M}$. This means that the state of Eq.~(\ref{eq:Psim-1}) represents a symmetric Dicke state where $(m-1)$ atoms are in the ground state and $N-(m-1)$ atoms are in the excited state.  The normalization factor $G_{\Phi_{N}}^{(m-1)}(\vec{r}_{1}, \hdots , \vec{r}_{1})$ in Eq.~(\ref{eq:rhom-1}) for $\theta_{1} = 0$ is thus equal to $\bra{\frac{N}{2},\frac{N}{2}} (J_{+})^{m-1}(J_{-})^{m-1} \ket{\frac{N}{2},\frac{N}{2}}=\binom{N}{m-1}((m-1)!)^{2}$, so that Eq.~(\ref{eq:GSdef}) can also be written as
\begin{equation}
\begin{aligned}
&G_{\Phi_{N}}^{(m)}(\vec{r}_{1}, \hdots , \vec{r}_{1},\vec{r}_{2}) \\
&= \bra{\Psi_{m-1}} E^{(-)}(\vec{r}_{2})E^{(+)}(\vec{r}_{2}) \ket{\Psi_{m-1}}  \binom{N}{m-1}((m-1)!)^{2} \\
&= I_{\Psi_{m-1}}(\vec{r}_{2}) \binom{N}{m-1}((m-1)!)^{2} \ ,
\end{aligned}
\end{equation}
with
\begin{equation}
I_{\Psi_{m-1}}(\vec{r}_{2}) = \bra{\Psi_{m-1}} E^{(-)}(\vec{r}_{2})E^{(+)}(\vec{r}_{2}) \ket{\Psi_{m-1}} \ .
\end{equation}
Note that $I_{\Psi_{m-1}}$ is the intensity of the radiation produced by atoms prepared in the Dicke state $\ket{\frac{N}{2},\frac{N}{2}-(m-1)}$, i.e. \cite{Wiegner(2011)PRA},
\begin{equation}
\begin{aligned}
& I_{\Psi_{m-1}}(\vec{r}_{2}) \\
& = (N - m + 1) \left( \frac{N-m}{N-1} + \frac{m-1}{N(N-1)}\frac{\sin^2(N\frac{\varphi_{12}}{2})}{\sin^2(\frac{\varphi_{12}}{2})}\right) \, .
\label{eq:IntensityPsi}
\end{aligned}
\end{equation}

We thus have the result that the superradiant character of the Dicke state $\ket{\frac{N}{2}, \frac{N}{2}-(m-1)}$ can be simulated by measuring the $m$-th order intensity correlation function $G_{\Phi_{N}}^{(m)}(\vec{r}_{1}, \hdots, \vec{r}_{1}, \vec{r}_{2})$ for $\theta_{1} = 0$ of the fully excited system:
\begin{equation}
\begin{aligned}
& {G}^{(m)}_{\Phi_{N}}(\vec{r}_1,..., \vec{r}_1, \vec{r}_2)  \\
&\hspace{2mm}=\frac{N!(m-1)!}{(N-m)!}\,\left( \frac{N-m}{N-1} + \frac{m-1}{N(N-1)}\frac{\sin^2(N\frac{\varphi_{12}}{2})}{\sin^2(\frac{\varphi_{12}}{2})}\right)\,.
\end{aligned}
\end{equation}
The first $(m-1)$ detections prepare the system conditionally in the Dicke state $\ket{\frac{N}{2}, \frac{N}{2}-(m-1)}$ whose behavior is probed by the detection of the $m$-th photon. Note from Eq.~(\ref{eq:IntensityPsi}) that the peak height is proportional to $N$ for $m=N$, and the width is proportional to $1/N$. These are the characteristic features of the Dicke superradiance when the Dicke state has one excitation.

Note that by putting $(m-1)$ detectors at $\theta_{1}=0$, we obtained symmetric Dicke states. However, for $\theta_{1} \neq 0$, we could get timed Dicke states as introduced by Scully and collaborators \cite{Scully(2006),*Scully(2007),*Scully(2008),*Scully(2009),*Scully(2009)Super}. For example a single detection will produce the state
\begin{equation}
\begin{aligned}
 W_{N,N-1}&= \frac{1}{\sqrt{N}} \sum_{l} e^{-i\, \varphi_{l1}} s_{-}^{(l)}\ket{\Phi_{N}}\\
& = \frac{1}{\sqrt{N}} \sum_{l} e^{-i\, \varphi_{l1}} \ket{e\hdots e\, g_{l} \hdots e}  \ ,
\end{aligned}
\end{equation}
in which one atom is in the ground state with a well defined phase factor.

\section{Compact way for calculating higher order intensity correlations}
\label{sec:functional}

In this section we finally present a compact form for calculating higher order intensity correlations by use of the characteristic functional. In this way different orders of the correlation functions can be obtained via simple differentiation. The characteristic functional $C[f(\cdot)]$ for normally ordered correlations can be defined via \cite{KLAUDER(1968)}
\begin{equation}
\begin{aligned}
	C[f(\cdot)] = &\left<\exp\left\{i\int f^{*}(\vec{r}) E^{(-)}(\vec{r})\text{d}^{3}r \right\}\right. \\
	&\hspace{10mm} \times \left.\exp\left\{i\int f(\vec{r}) E^{(+)}(\vec{r})\text{d}^{3}r \right\} \right> \ .
\label{eq:FunctionalDef}
\end{aligned}
\end{equation}
For $N$ sources we will use the discrete version
\begin{equation}
	E^{(+)}(\vec{r}_{l})=\left[E^{(-)}(\vec{r}_{l})\right]^{\dagger}= \sum_{j=1}^{N}{s}_{-}^{(j)}c_{l , j} \ ,
\end{equation}
where $c_{l,j}=\exp\left\{-i\frac{\omega}{c}\, \vec{n}_{l}\cdot \vec{R}_{j}\right\}$ describes the accumulated phase of a photon traveling from source $j$ to detector $l$ (cf. Eq.~\ref{eq:field}). In this way we obtain
\begin{equation}
\begin{aligned}
	C&[f(\cdot)]\\
	  &=\left<\exp\left[i\sum_{l=1}^{m}f^{*}_{l}E^{(-)}(\vec{r}_{l})\right] \exp\left[i\sum_{l=1}^{m}f_{l}E^{(+)}(\vec{r}_{l})\right] \right> \ ,
\label{eq:Cdot2}
\end{aligned}
\end{equation}
where $f_{l}$ is the discrete version of $f(\vec{r})$.
Hereby, the sum over the electric field components can also be written in the form
\begin{equation}
	\sum_{l=1}^{m}f_{l}E^{(+)}(\vec{r}_{l})=\sum_{l=1}^{m}\sum_{j=1}^{N}{s}_{-}^{(j)}c_{l , j}f_{l}=\sum_{j=1}^{N}{s}_{-}^{(j)}\beta_{j} \ ,
\label{eq:sumElectric}
\end{equation}
where $\beta_{j}$ is defined as
\begin{equation}
\beta_{j}=\sum_{l=1}^{m}c_{l , j}f_{l} \ .
\label{eq:definitionbeta}
\end{equation}
In general, to derive the $m$-th order correlation function $G^{(m)}(\vec{r}_{1},\vec{r}_{2},\hdots,\vec{r}_{m})$ one can utilize the functional differentiation of the characteristic functional (cf. Eq.~(\ref{eq:Cdot2}))
\begin{equation}
\begin{aligned}
	&G^{(m)}(\vec{r}_{1},\vec{r}_{2},\hdots,\vec{r}_{m}) \\
	&\hspace{5mm}= \left. (-1)^{m} \frac{\partial^{2m}C[f(\cdot)]}{\partial f_{1}^{*}\hdots \partial f_{m}^{*}\partial f_{1}\hdots \partial f_{m}} \right|_{\substack{f_{1}=\hdots=f_{m}=0 \\
	f^{*}_{1}=\hdots=f^{*}_{m}=0}} \ .
\label{eq:DefGmSPE}
\end{aligned}	
\end{equation}
Since all atoms are independent of each other, Eq.~(\ref{eq:Cdot2}) can be factorized and therefore changed into a product-form
\begin{equation}
\begin{aligned}
	C[f(\cdot)]=& \prod_{j=1}^{N} \left< e^{i{s}_{+}^{(j)}\beta_{j}^{*}} e^{i{s}_{-}^{(j)}\beta_{j}} \right> \\
	=& \prod_{j=1}^{N} \left< \left(1 + i{s}_{+}^{(j)}\beta_{j}^{*}\right) \left(1 + i{s}_{-}^{(j)}\beta_{j}\right) \right> \\
	=& \prod_{j=1}^{N} \left( 1 - |\beta_{j}|^{2} \right) = \prod_{j=1}^{N} \left( 1 - \big| \sum_{l=1}^{m} c_{l , j}f_{l} \big|^{2} \right)  \ ,
\label{eq:definitionCdot}	
\end{aligned}
\end{equation}
where we exploited the fact that all atoms are fully excited. We have thus obtained a compact form for the characteristic functional for the radiation from a sample of fully excited atoms.
Again, considering the superradiant case by choosing
\begin{equation}
\begin{aligned}
f_{i}&=f_{1} \ (i=1,2,\hdots,m-1) \\
f_{m}&=f_{2} \ ,
\end{aligned}
\end{equation}
we can calculate the explicit form of the correlation functions from Eq.~(\ref{eq:definitionCdot}). This procedure leads to Eq.~(\ref{GMGENSbb}).

\section{Conclusion}
\label{sec:conclusion}

The foregoing theoretical investigations show that beyond entangled symmetric Dicke states it is also possible to employ statistically independent and initially uncorrelated incoherent light sources to obtain a focused spatial emission pattern of the emitted radiation. For $N$ initially uncorrelated two-level atoms the directional spontaneous emission of the $m$-th photon is due to preceding measurements of $(m-1)$ photons along selected directions, projecting the uncorrelated atoms into Dicke states $\ket{\frac{N}{2},\frac{N}{2}-(m-1)}$ with $N-(m-1)$ excitations ($N \geq m > 1$). Our work thus affirms the great importance of projective measurements and demonstrates how higher order intensity intensity correlations could be used to implement different physical phenomena.

\section{Acknowledgement}

The authors gratefully acknowledge funding by the Erlangen Graduate School in Advanced Optical Technologies (SAOT) by the German Research Foundation (DFG) in the framework of the German excellence initiative. R.W. and S. O. gratefully acknowledge financial support by the Elite Network of Bavaria and the hospitality at the Oklahoma State University. This work was supported by the DFG research grant ZA 293/4-1. G.S.A. is especially grateful to the SAOT for providing financial grant for making this collaboration possible.

\bibliography{literatur_superradiance}
\bibliographystyle{apsrev4-1}

\end{document}